\documentclass[prb,twocolumn,aps,showpacs,fixfloats]{revtex4}
\usepackage{graphicx}
\usepackage{bm}
\usepackage{amsmath,amssymb}
\usepackage{subfigure}
\usepackage{float}
\usepackage{latexsym}
\usepackage{color}
\usepackage{enumerate}
\usepackage{pdfpages}
\usepackage{tikz}
\usepackage{hyperref}

\begin{document}
\newcommand{\s}{\scriptscriptstyle}
\newcommand{\uu}{\uparrow \uparrow}
\newcommand{\ud}{\uparrow \downarrow}
\newcommand{\du}{\downarrow \uparrow}
\newcommand{\dd}{\downarrow \downarrow}
\newcommand{\ket}[1] { \left|{#1}\right> }
\newcommand{\bra}[1] { \left<{#1}\right| }
\newcommand{\bracket}[2] {\left< \left. {#1} \right| {#2} \right>}
\newcommand{\vc}[1] {\ensuremath {\bm {#1}}}
\newcommand{\tr}{\text{Tr}}
\newcommand{\Trans}{\ensuremath \Upsilon}
\newcommand{\Refl}{\ensuremath \mathcal{R}}

\title{Effective tunnel conductance and effective ac conductivity
of randomly strained graphene}

\author{Rajesh K. Malla   and M. E. Raikh}

\affiliation{ Department of Physics and
Astronomy, University of Utah, Salt Lake City, UT 84112}

\begin{abstract}
We consider a single-layer  graphene with high ripples, so that the pseudo-magnetic fields due
to these ripples are strong. If the magnetic length corresponding to a typical pseudo-magnetic field
is smaller than the ripple size, the resulting Landau levels are local.
Then the effective properties of the macroscopic sample can be calculated
by averaging the local properties over the distribution of ripples. We find that this averaging does not wash out the Landau quantization completely. Average density of states (DOS) contains a feature (inflection point) at energy corresponding to the first Landau level in a {\em typical} field. Moreover, the frequency dependence of the ac conductivity
%while the average ac conductivity
contains a maximum at a frequency corresponding to the first Landau level in a typical field. This nontrivial behavior of the effective characteristics of randomly strained graphene is a consequence of non-equidistance of the Landau levels in the Dirac spectrum.
\end{abstract}

\maketitle

\section{Introduction}

It is known that graphene deposited on different substrates,
such as SiO$_2$, exhibits ripples.\cite{deposition1} These ripples
essentially follow the corrugation of the substrate.

Already in the early days of the research on graphene
it was predicted\cite{deposition2,deposition2'} (for review see
Refs. \onlinecite{deposition3,Maria1}) that  random strain,
caused by ripples, can play the role of random magnetic field
affecting the orbital motion of electrons and having opposite signs
in different valleys.

Later, a direct evidence of the substrate-induced ripples was reported in
Refs. \onlinecite{HeinzSTMimage,FuhrerTopography1,TopographySiO2,FuhrerTopographySi02,Zhitenev},
where the topography of the surface was revealed via
the STM (scanning tunneling microscopy) images.
The fact that the ripples indeed give rise
to the pseudo-magnetic fields was directly confirmed
in the experiments Refs. \onlinecite{BerkleyBubble,MosesChan,AnotherTunneling}.
In Refs. \onlinecite{BerkleyBubble,MosesChan,AnotherTunneling}
the staircases of the Landau levels were observed
in the bias dependence of the local tunnel conductance.
Another physical manifestation of the curvature-induced magnetic field
is the sensitivity of transport to the external magnetic field applied parallel to
the surface. This sensitivity was demonstrated
experimentally in Refs. \onlinecite{Folk1,Folk2}.
Due to strain, a parallel magnetic field generates
a pseudo-magnetic field
normal to the surface which causes the electron
dephasing.\cite{Maria,Lewenkopf,Lewenkopf1}
%An
Alternative explanation\cite{Baranger} is common for electron motion along the rough interface:
%The underlying reason is that,
in course of this motion
%along the rippled surface,
electron trajectory can enclose a finite flux even if the field is parallel to the graphene sheet
{\em on average}.

%A quantitative relation between the arbitrary surface profile, $h(x,y)$, and the effective perpendicular magnetic field created
% due to ripples in the applied parallel magnetic field was explored in Refs.
% \onlinecite{Lewenkopf,Lewenkopf1}.
%\begin{figure}
%\includegraphics[scale=0.3]{cartoon.pdf}
%\includegraphics[scale=0.65]{cartoon1.pdf}
%\caption{(Color online) (a) Schematic illustration of random pseudo-magnetic field
%induced by strain. In the red and blue regions the field has opposite signs.
%Typical size, $R_c$, of the regions is assumed to be much bigger than typical magnetic length. Low-energy states originate from the white regions where magnetic field passes through zero. For these ``snake"-like states electron propagates along the boundary $B=0$ and decays into both red and blue regions. (b) Since the extension, $\delta y$, of the electron wave-function in the $y$-direction, which is normal to the direction of propagation, is much smaller than $R_c$, the magnetic field within
%$\delta y$ can be linearized.}
%\label{fcartoon}
%\end{figure}

The motivation for the current studies of the strained graphene\cite{Experiment,Andrei}
is a promise to manipulate the electron spectrum by tailoring the strain.\cite{Experiment1}
For a review on
``strain engineering" see
%e.g.
Ref. \onlinecite{FengLiuReview}.
\begin{figure}[h!]
\includegraphics[scale=0.3]{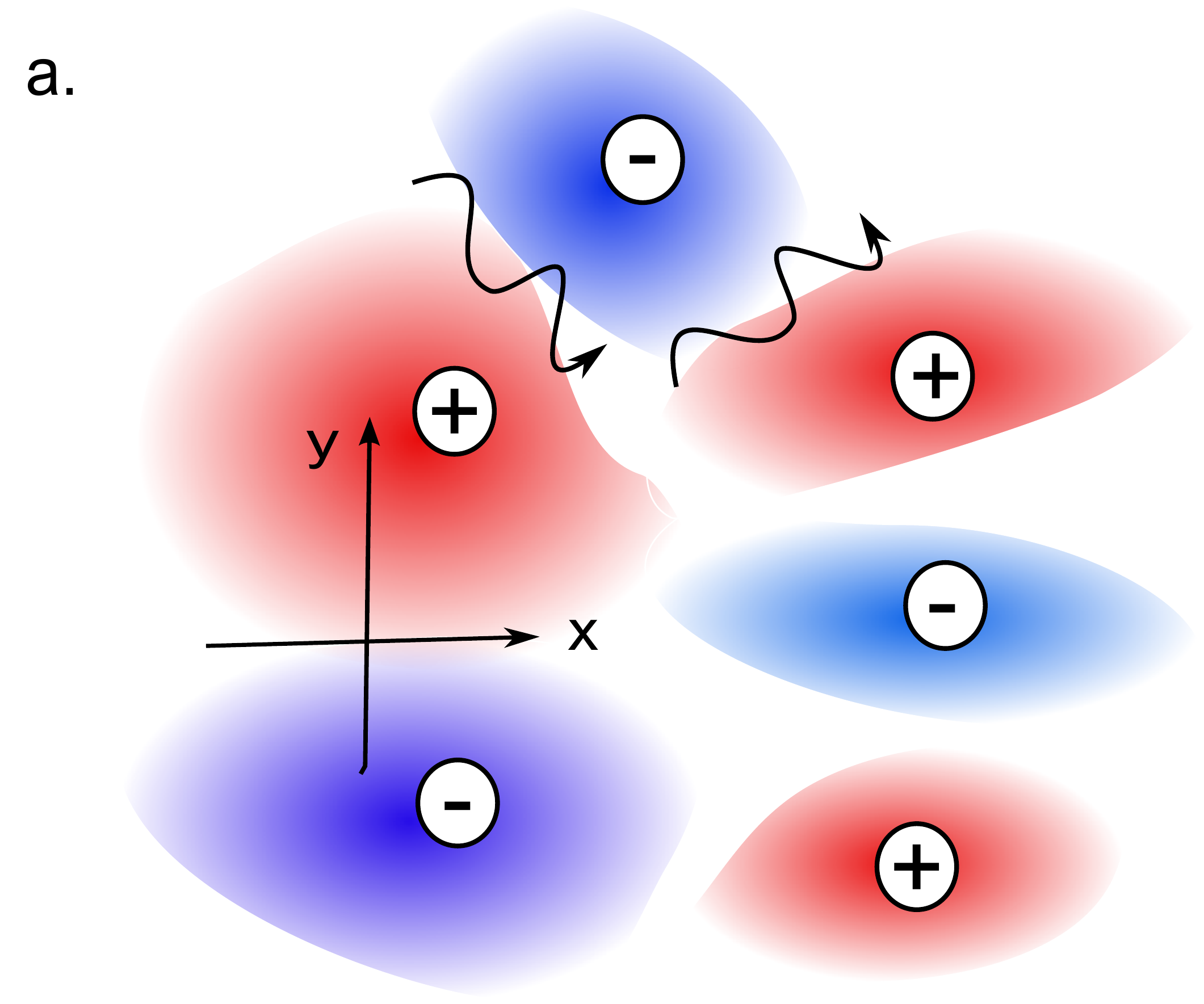}
\includegraphics[scale=0.65]{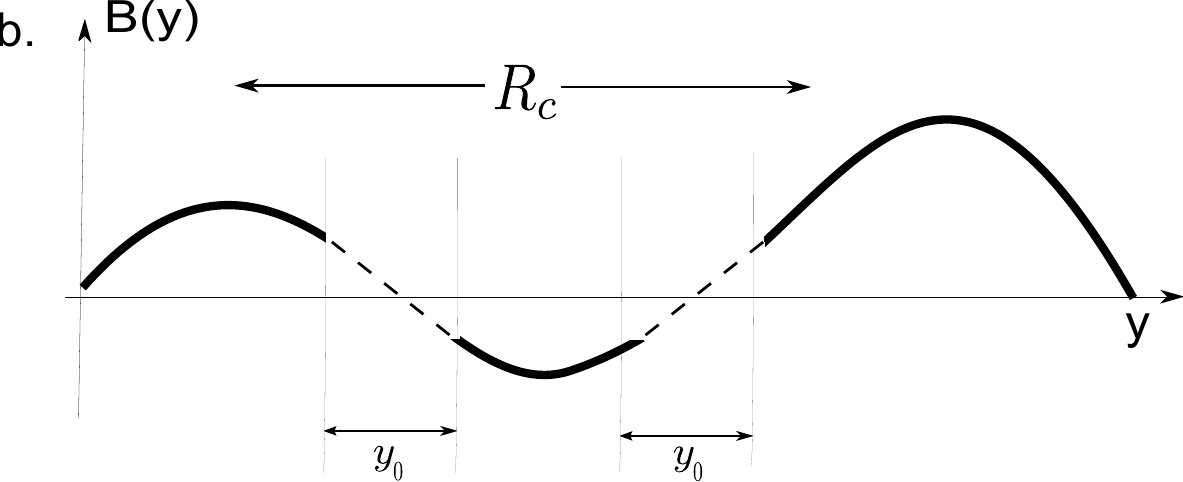}
\caption{(Color online) (a) Schematic illustration of random pseudo-magnetic field
induced by strain. In the red and blue regions the field has opposite signs.
Typical size, $R_c$, of the regions is assumed to be much bigger than the typical magnetic length. Low-energy states originate from the white regions where magnetic field passes through zero. For these ``snake"-like states electron propagates along the boundary $B=0$ and decays into both red and blue regions. (b) Since the extension, $y_0$, of the electron wave-function in the $y$-direction, which is normal to the direction of propagation, is much smaller than $R_c$, the magnetic field within
$y_0$ can be linearized.}
\label{fcartoon}
\end{figure}
In the present paper we address one peculiar property of graphene
with pseudo-magnetic fields. These fields create position-dependent staircases of
Landau levels. Randomness in the ``steps" of these staircases tends to smear the
average density of states (DOS).
It is not obvious whether or not this smearing is complete.
%Recall that, for a semiconductor with a parabolic spectrum, the effect of a random field is eliminated completely from the average DOS.
Specifics of graphene with linear dispersion spectrum is that Landau levels are not equidistant in a constant magnetic field. We demonstrate that, as a consequence of this
non-equidistance, the average DOS {\em retains the memory} about the
Landau quantization.

%{\em even after averaging}.

If the random field changes smoothly along the plane,
local Landau levels are formed in the strained regions.
They manifest themselves in the local tunnel
conductance.\cite{BerkleyBubble,MosesChan}
We demonstrate that the average (over the surface area) tunnel conductance, $G$, still contains a feature reflecting the first Landau level in a {\em typical} field. Naturally, this feature is more pronounced in the derivative, $\frac{dG}{dV}$, of the conductance with respect to the bias voltage, $V$. This derivative saturates at high bias.
%$V$.
Together with the average tunnel conductance, we calculate the average frequency-dependent ac conductivity, $\sigma(\Omega)$. It appears that a feature at $\Omega$
corresponding to the first Landau level in a typical field not only survives the
averaging, but manifests itself stronger than in the tunnel conductance.

The bigger is the correlation length, $R_c$, of the random magnetic field, see
Fig. \ref{fcartoon}, the more pronounced is the Landau quantization in the
average DOS, or, in other words, the stronger is the
depletion of the DOS near zero energy.
Finite $R_c$ smears the quantization due to the  emergence of the low-energy
states. These states, which we study in the present paper, originate from the spatial regions where the random magnetic field passes through zero, as illustrated in Fig.~\ref{fcartoon}.

\section{Average DOS}
If the magnetic field does not change in space, then the DOS is given by
\begin{equation}
\label{DOSinB}
\rho(E,B) = \frac{1}{2\pi \ell_B^2}\sum_{n=-\infty}^{\infty}\delta\left(E - E_n\right),
\end{equation}
where the energies, $E_n$, are the Landau levels in graphene
\begin{equation}
E_n={\text sgn}(n)\left(2e\hbar v_F^2|n|B\right)^{1/2}.
\end{equation}
Here $v_F\approx 10^6$m/s is the Fermi velocity, and
\begin{equation}
\ell_B=\left(\frac{\hbar}{eB}\right)^{1/2}
\end{equation}
is the magnetic length.

Actual shape, $\langle \rho(E,B)\rangle$, of the average DOS  depends on the relation between the correlation radius, $R_c$, of the random field and the magnetic length, $\ell_B$, in the characteristic field, $B_0$.
For $R_c\gg \ell_B$ adiabatic description applies. Within this description, the
local DOS is given by Eq. (\ref{DOSinB}) at every point, with
$E_n$ being the levels in the field, $B$, at this point.
Then $\langle \rho(E)\rangle$ is obtained by averaging of
Eq. (\ref{DOSinB}) over the distribution of $B$, i.e.
 \begin{equation}
\label{DOSinB1}
\langle \rho(E)\rangle=\frac{1}{B_0}\int\limits_{-\infty}^{\infty}dB \rho(E,B)F\Big(\frac{B}{B_0}\Big),
\end{equation}
where the distribution function, $F$, is normalized, so that
$\int_{-\infty}^{\infty}dx F(x)=1$.

The integration in Eq. (\ref{DOSinB1}) is performed with the help of
the $\delta$-functions. Prior to the averaging, it is convenient to
rewrite the DOS Eq. (\ref{DOSinB}) in the form
\begin{equation}
\label{averaged}
\rho(E,B)=\frac{e|B|}{2\pi\hbar}\delta(E)+\frac{|E|}{\hbar^2 v_F^2}\sum_{n=1}^{\infty}
\frac{|B|}{n}\delta\left(|B|-\frac{E^2}{2\pi e\hbar v_F^2 n}   \right).
\end{equation}

\begin{figure}
\includegraphics[scale=0.45]{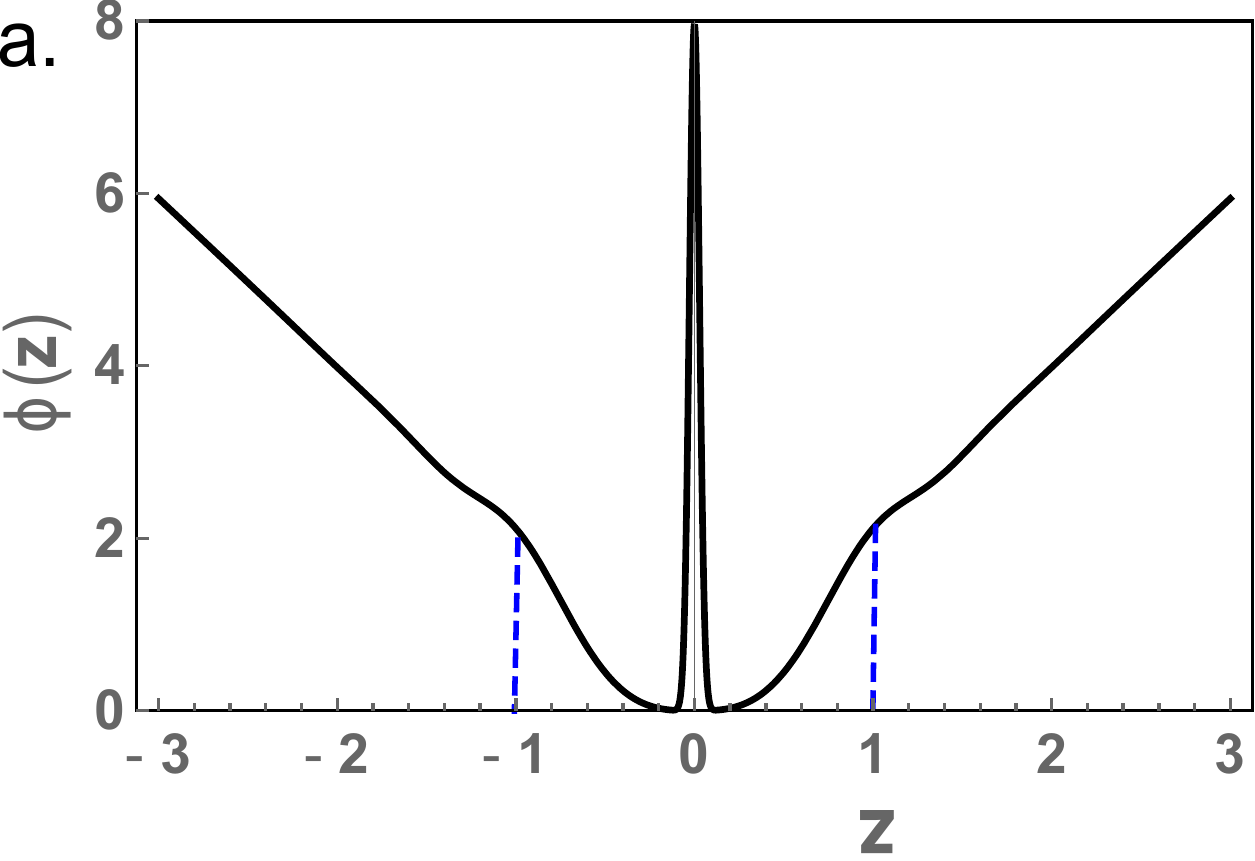}
\includegraphics[scale=0.47]{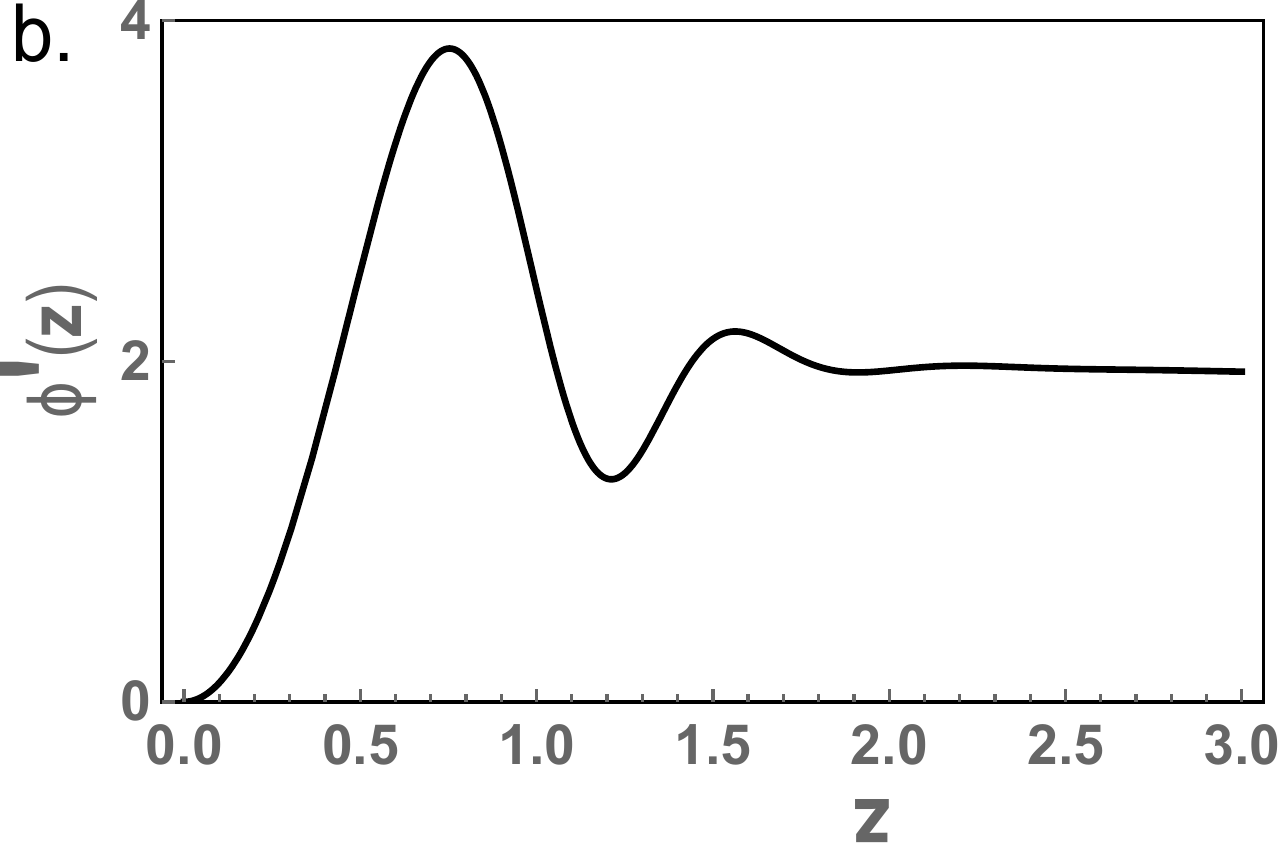}
\caption{(Color online) (a) Average DOS in randomly
strained graphene is plotted from Eq. (\ref{averaged}) versus the dimensionless
energy $z=E/w$, where $w$ is the Landau quantization energy
in a {\em typical} pseudo-magnetic field. For demonstration purpose the $\delta$-peak at $E=0$ is broadened into a gaussian with a width $0.04$. Note
that the first Landau level in a {\em typical} field survives the averaging
and manifests itself as a feature near $z=1$. To elucidate this feature the
derivative of average DOS is plotted in (b). It reveals
not only the first but also the second Landau level.  }
\label{fDOS1}
\end{figure}
Obviously, the $n=0$ Landau level
is not smeared by averaging. However, the magnitude of the $\delta(E)$-term
contains the magnetic field via $\ell_B$ and, thus, should be averaged.
Equally, in $n\neq 0$ terms, the values of $B$ at which the arguments
of the $\delta$-functions turn to zero, should be substituted into the
prefactor $\frac{1}{\ell_B^2}$.

It follows from Eq. (\ref{DOSinB1}) that the characteristic scale of change of the
average DOS is related to $B_0$ as
\begin{equation}
\label{scale}
w=\Bigg(\frac{2ev_{\s F}^2B_0}{\hbar }\Bigg)^{1/2},
\end{equation}
so that $\langle \rho(E)\rangle$ is a function of the dimensionless ratio
\begin{equation}
\label{ratio}
z=\frac{E}{w}.
\end{equation}
We cast the result of averaging into the form
\begin{equation}
\langle \rho(E)\rangle=\frac{1}{2\pi \ell_{B_0}^2w}\phi\left(\frac{E}{w}\right).
\end{equation}
where the dimensionless function $\phi(z)$ is given by
\begin{equation}
\label{form}
\phi(z)=2\overline{F}\delta(z)+4|z|^3\sum_{n=1}^{\infty}\frac{1}{n^2}F\Big(\frac{z^2}{n} \Big),
\end{equation}
where $\overline{F}=\int_0^{\infty}dz zF(z)$.

For Gaussian $F(x)=\frac{1}{\pi^{1/2}}\exp(-x^2)$
we have ${\overline F}=\frac{1}{2\pi^{1/2}}$. The small-$z$
asymptote of $\phi(z)$ follows from Eq. (\ref{form})
upon setting $F(z)=\frac{1}{\pi^{1/2}}$
\begin{equation}
\label{small-z}
\phi(z)|_{|z|\ll 1} \approx \frac{4|z|^3}{\pi^{1/2}}\sum_{n=1}^{\infty}\frac{1}{n^2}=\frac{2\pi^{3/2}}{3}|z|^3.
\end{equation}
This behavior reflects the depletion of the average DOS at small energies due to the formation of the Landau levels.
The large-$z$ asymptote,
$\phi(z)\approx 2|z|$,
is determined by the terms $n\sim z^2 \gg 1$,
when the summation can be replaced by integration.
Naturally, it recovers the DOS in a zero field.
Our main finding is that, at intermediate $z$,
the function $\phi(z)$ exhibits a feature (inflection point) near
$z=1$ as shown in Fig. \ref{fDOS1}a.
Presence of this feature indicates that the
first Landau level {\em survives the averaging}. The
feature becomes even more pronounced in the derivative $\phi'(z)$ plotted in Fig.  \ref{fDOS1}b. Actually, as seen in Fig.  \ref{fDOS1}b, the derivative $\phi'(z)$
%Fig.  \ref{fDOS1}(b)
contains an additional feature which corresponds to the second Landau level,
so that it also survives the averaging in some form.

\section{Average ac conductivity}

A standard expression for the ac conductivity, $\sigma (\Omega)$, in a constant magnetic field, see e.g. Ref. \onlinecite{Magnetooptic}, reads
\begin{multline}
\label{Sigma}
\sigma(\Omega)=\frac{e^2v_F^2|eB|}{ \Omega}\sum_{n=0}^{\infty}
\delta\Big(|E_n|+|E_{n+1}|-\Omega\Big)\\
{\small \times}\Big\{\big[n_F(-|E_n|)-n_F(|E_{n+1}|)\big] + \big[n_F(-|E_{n+1}|)-n_F(|E_{n}|)\big]\Big\}.\hspace{-4mm}
\end{multline}
Here $n_F(E)=\left[\exp(E/T)+1\right]^{-1}$ is the Fermi distribution.
The first term in the square brackets describes the transitions from the occupied hole level $(n+1)$
to the empty electron level $n$, while the second term describes the transitions from the occupied hole level $n$
to the empty electron level $(n+1)$.

In the limit $\Omega \gg w$, Eq. (\ref{Sigma}) yields $\sigma=\frac{e^2}{4\hbar}$,
which is a standard result.
Averaging over the distribution of pseudo-magnetic fields in
Eq. (\ref{Sigma}) is quite similar to the averaging of the DOS.
The result is expressed through a dimensionless function,
$\Upsilon({\tilde z})$, of the dimensionless frequency ${\tilde z}=\Omega/w$.
This result reads
\begin{equation}
\sigma(\Omega)=\frac{e^2}{4\hbar}\Upsilon({\tilde z}).
\end{equation}
The form of the function $\Upsilon({\tilde z})$ is the following
\begin{equation}
\label{ACaverage}
\Upsilon({\tilde z})=8{\tilde z}^2\sum_{n=0}^{\infty}\frac{f_n({\tilde z},T)}{\left(\sqrt{n}+\sqrt{n+1}\right)^4}
F\Bigg(\frac{{\tilde z}^2}{\left(\sqrt{n}+\sqrt{n+1}\right)^2}\Bigg),
\end{equation}
where the temperature factor, $f_n({\tilde z},T)$,     is defined as
\begin{equation}
\label{Temperature}
f_n({\tilde z},T)=\frac{\exp(\lambda_T{\tilde z})}{4\cosh\left(\frac{\lambda_T{\tilde z}\sqrt{n}}{\sqrt{n}+\sqrt{n+1}}   \right)\cosh\left(\frac{\lambda_T{\tilde z}\sqrt{n+1}}{\sqrt{n}+\sqrt{n+1}}   \right)},
\end{equation}
with $\lambda_T=\frac{w}{2T}$.

\begin{figure}
\includegraphics[scale=0.5]{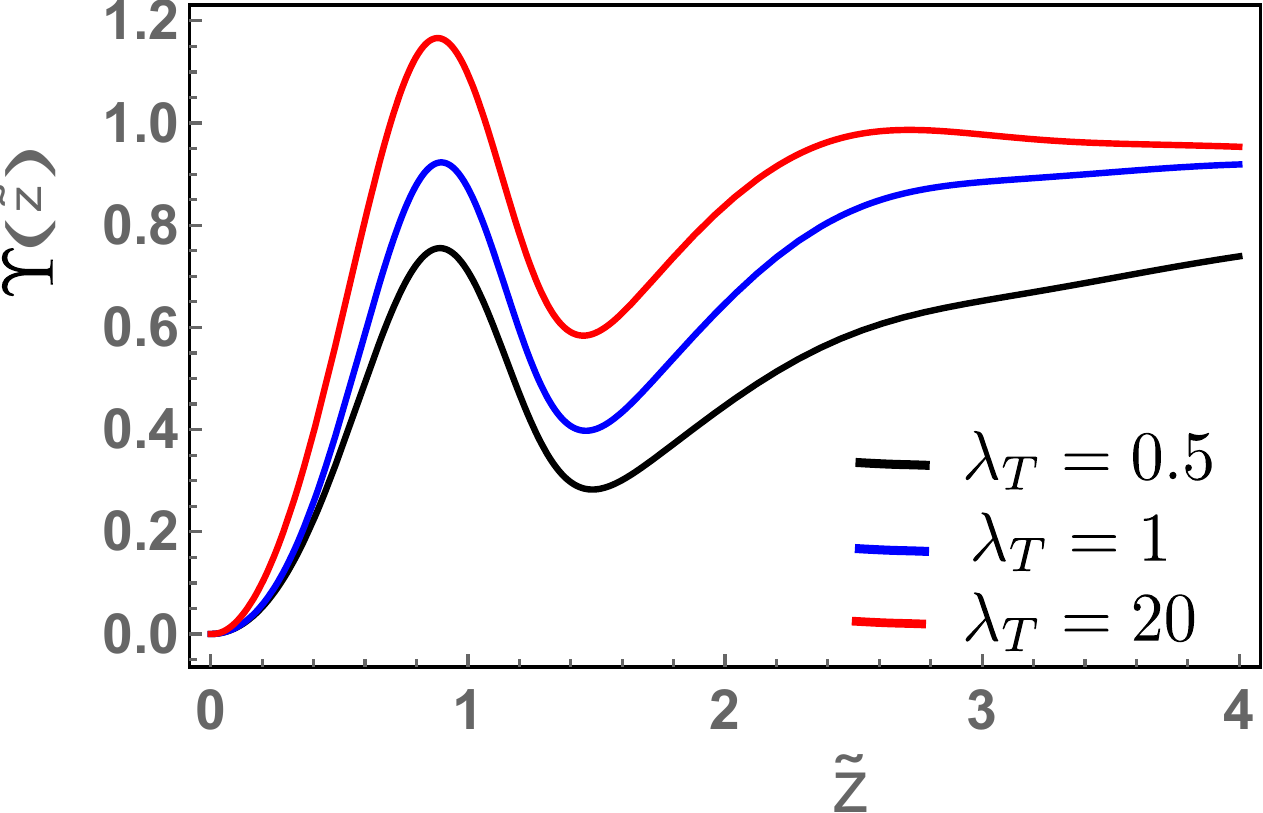}
\caption{(Color online) The average ac conductivity
is plotted from Eq. \ref{ACaverage} versus the dimensionless frequency ${\tilde z}=\Omega/w$,
where the characteristic frequency, $w$, is defined by Eq. (\ref{scale}).
Unlike the DOS, the feature corresponding to the transition $0\rightarrow 1$ emerges even without taking the derivative. Black, blue, and red curves correspond to the temperatures $T=w$,
$T=\frac{w}{2}$, and $T=\frac{w}{40}$, respectively.  }
\label{fconductivity}
\end{figure}

Note that there is an important difference between Eqs.~(\ref{form}) and (\ref{ACaverage}). Namely,
the summation in $\Upsilon({\tilde z})$ includes the term $n=0$. Due to this
term, a peak in the average ac conductivity is present even without taking derivative, as shown in Fig. \ref{fconductivity}. This peak corresponds to
the virtual transitions $0\rightarrow 1$ and $-1\rightarrow 0$ in the Kubo
formula. It is seen from Fig. \ref{fconductivity} that the peak remains pronounced
up to high temperatures.

\section{Finite-$R_c$ correction}
As it was mentioned above, the adiabatic description applies when the correlation length $R_c$ exceeds the magnetic length in a typical field. Even when this
condition is met, there are low-energy states in a random magnetic field for which adiabatic description
is inapplicable. These states originate from the regions where the random field
passes through zero. As illustrated in Fig. \ref{fcartoon}a, the contours $B=0$ form a percolation network. As a result of the smallness of $B$, the corresponding local magnetic length is big and exceeds $R_c$. To establish the portion of these
regions and their contribution to the DOS a full quantum mechanical
description is required. The simplification which allows one to develop this
description comes from the fact that the motion along the network $B=0$ has
a one-dimensional character.\cite{Chalker1994}

We start from the effective-mass Hamiltonian for an electron in the presence of an arbitrary magnetic field
\begin{equation}
\label{Dirac}
\hat{H}=v_F \begin{pmatrix}
0 && \hat{\pi}_x-i\hat{\pi}_y \\ \\ \hat{\pi}_x+i\hat{\pi}_y && 0
\end{pmatrix},
\end{equation}
where  $\hat{\pi}={\bf \hat{k}}+\frac{e}{c}{\bf A}$, with ${\bf \hat{k}}$ being the momentum operator and ${\bf A}$ being the vector potential. The Hamiltonian Eq.~(\ref{Dirac}) describes the states in a single valley.
% written for
%one valley.
\begin{figure}
\includegraphics[scale=0.3]{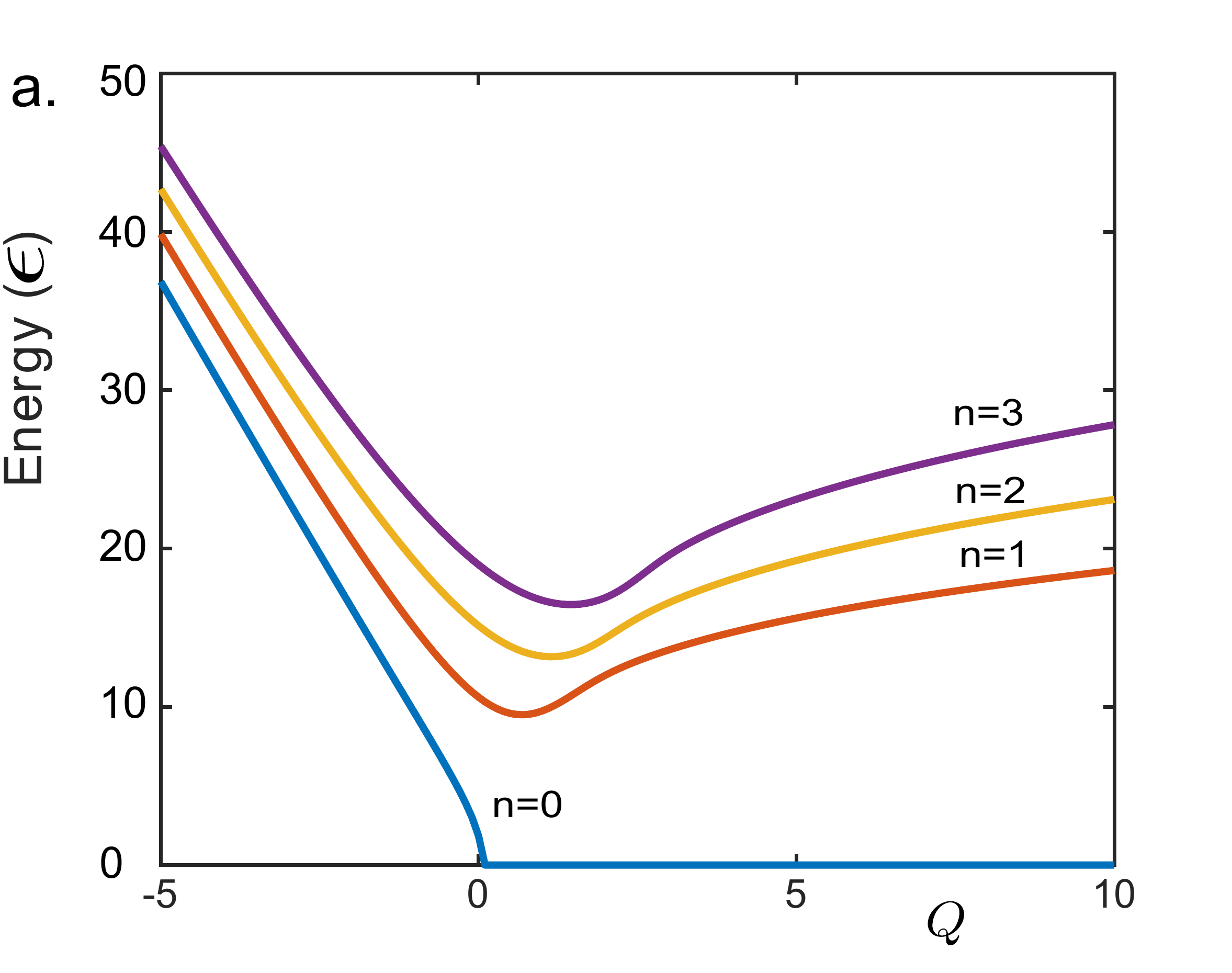}
\includegraphics[scale=0.32]{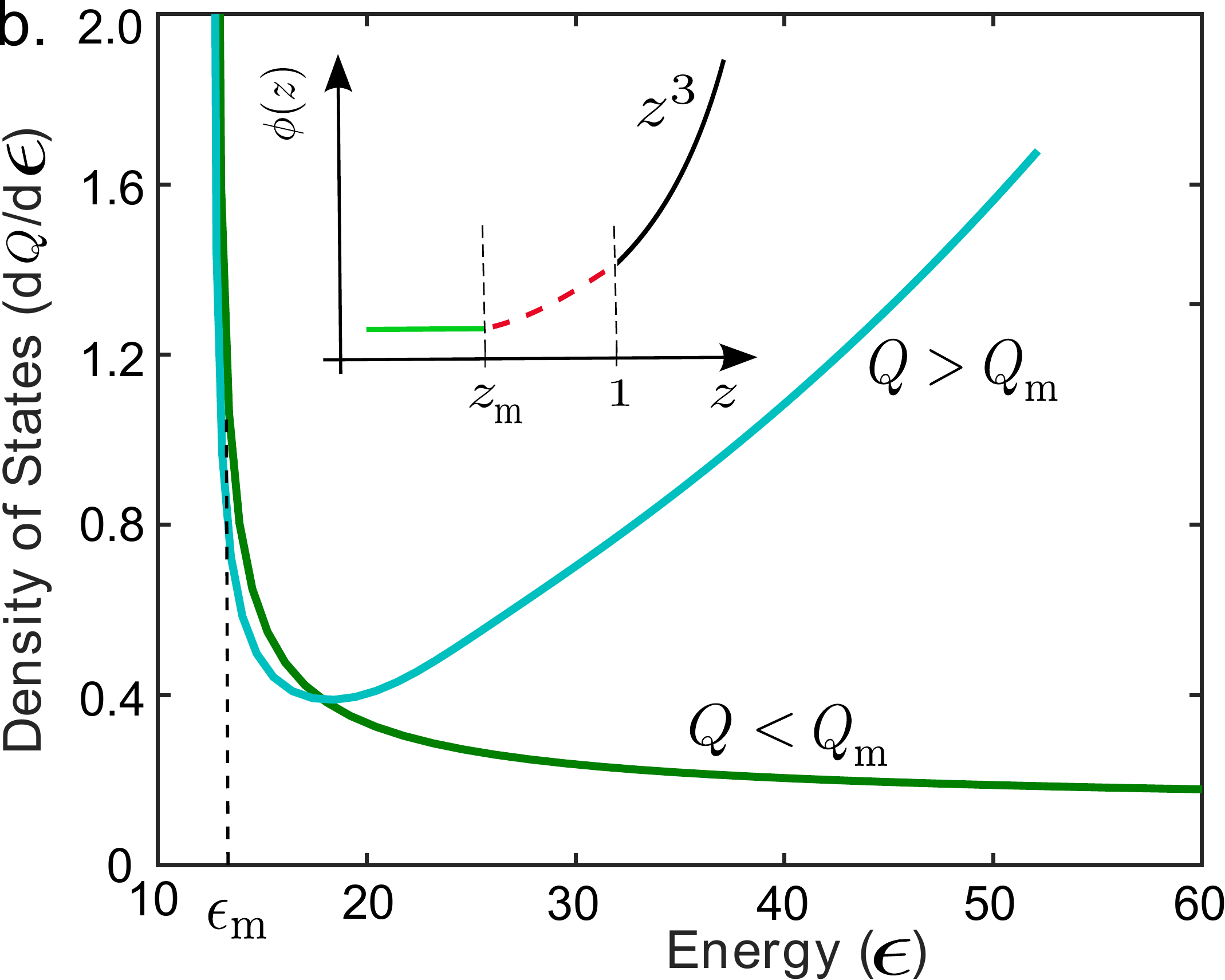}
\caption{(Color online) (a) First four branches of the dispersion of the snake states
obtained by the numerical solution of Eq.~(\ref{Schrodinger}).
(b) Two contributions to the dimensionless DOS, $\frac{dQ}{d\epsilon}$, are
plotted from (a) for the branch $n=1$. Inset: schematic behavior of the dimensionless DOS plotted in
Fig. \ref{fDOS1} in the low-energy domain. The dependence $\phi(z)\propto z^3$ extends down to
$z=z_m=\left(\frac{\ell_B}{R_c}\right)^{1/3}$ and is terminated as $z<z_m$, where the main contribution
to the DOS comes from the snake states. }
\label{fenergyplot}
\end{figure}
The crucial assumption, which will be justified later, is that, near the contour $B=0$, see Fig. \ref{fcartoon}a, the magnetic field can be linearized, see Fig. \ref{fcartoon}b, as
$B=\frac{2B_0y}{R_c}$, so that the vector potential in the Landau gauge is given by ${\bf A}=(-\frac{B_0y^2}{R_c},0)$. This form of the vector potential allows the
separation of variables in the Dirac equation, namely
\begin{equation}
\label{Wavefunction}
\Psi=e^{i k_x x}\begin{pmatrix}
\psi_1(y) \\ \\ \psi_2(y)
\end{pmatrix},
\end{equation}
where the components $\psi_1(y)$ and $\psi_2(y)$ satisfy the system of equations
\begin{eqnarray}
\label{Dirac1}
\frac{E}{v_F}\psi_1=\left(k_x -\frac{e B_0 y^2}{c R_c}\right)\psi_2+ \frac{d \psi_2(y)}{d y},\nonumber \\
\frac{E}{v_F}\psi_2=\left(k_x -\frac{e B_0 y^2}{2 R_c}\right)\psi_1-\frac{d \psi_1(y)}{d y}.
\end{eqnarray}
We can now introduce the dimensionless variables
\begin{equation}
\label{units}
y=y_0 s, ~~k_x=\frac{Q}{y_0},~~
\epsilon= y_0\frac{E}{v_F},
\end{equation}
where $y_0$ is given by
\begin{equation}
\label{y0}
y_0=\left(\ell_B^2 R_c\right)^{1/3},
\end{equation}
and reduce the system Eq. (\ref{Dirac1}) to a single second-order differential equation for $\psi_2$
\begin{equation}
\label{Schrodinger}
\frac{d^2 \psi_2}{d s^2}+\Big[\left(Q-s^2 \right)^2 +2s\Big]\psi_2=\epsilon^2 \psi_2.
\end{equation}
This equation has the form of the Schr{\"o}dinger equation with potential $U(s)=\left[\left(Q-s^2 \right)^2 +2s\right]$. Note that the term $2s$ in the potential is specific for the Dirac electron. For the Schr{\"o}dinger electron in a magnetic field\cite{Chalker1994} $B(y)\propto y$
this term is absent. It is this term which ensures
that the level $\epsilon=0$ is robust.

From the expression Eq. (\ref{y0}) for $y_0$ we conclude
that, by virtue of the condition $R_c \gg \ell_B$, the length $y_0$ lies in the interval
\begin{equation}
\label{condition}
R_c \gg  y_0 \gg \ell_B.
\end{equation}
First relation justifies linearizing of the magnetic
field performed above. Second relation indicates that the
energies of the snake states, described by the system
Eq. (\ref{Dirac1}), are much smaller than the characteristic energy, $w$.

Consider a ripple of a size $\sim R_c$. Its area is $\sim R_c^2$. Snake states
are confined along the perimeter, within a strip of a width $\sim y_0$, see Fig. \ref{fcartoon}. The area of the strip is $\sim y_0R_c$. Thus,
for energies of the order of $w$, the relative correction,
$\rho_{sn}(w)$, to the DOS due to finite $R_c$ is of the order of
\begin{equation}
\label{correction}
\frac{\rho_{sn}(w)}{\langle \rho (w)\rangle}\sim
\frac{y_0}{R_c}\sim\left(\frac{\ell_B}{R_c}\right)^{1/3}.
\end{equation}

We now  turn to the low-energy domain $E \ll w$,
where the DOS is dominated by the
snake states.  The dispersion law, $\epsilon(Q)$,
of the snake states, drifting along the contour $B=0$,
can be found analytically in the limits of large
negative and large positive $Q$. For negative $Q$,
such that $|Q|\gg 1$, the potential can be replaced
by $U(s)=Q^2+2|Q|s^2$. Then the Schr{\"o}dinger equation
Eq. (\ref{Schrodinger}) reduces to the harmonic
oscillator equation yielding
\begin{equation}
\label{asymptote1}
\epsilon_n^2=Q^2+(2n+1)|2Q|^{1/2},~~~~~ Q<0,~~ |Q|\gg 1.
\end{equation}
At large positive $Q$ the potential
should be expanded near $s=Q^{1/2}$.
Upon setting $s=\pm Q^{1/2}+s_1$,
and assuming that $s_1\ll Q^{1/2}$,
the potential takes the form $U(s)\approx 2Q^{1/2}+4Qs_1^2$,
which is again the harmonic oscillator
potential with frequency $2Q^{1/2}$.
This leads to the following
dispersion laws of the branches
\begin{equation}
\label{asymptote2}
\epsilon_n^2=4nQ^{1/2},~~~~~ Q\gg 1.
\end{equation}
Integer $n$ in Eqs. (\ref{asymptote1}),
(\ref{asymptote2}) starts from $n=0$.
Dispersion laws of the first four branches calculated numerically
are shown in Fig. \ref{fenergyplot}a.
The asymptotes Eqs. (\ref{asymptote1}), (\ref{asymptote2}) are clearly seen in the plots.
At large positive $Q$ the branches approach
asymptotically the Landau levels in the bulk. The branch $n=0$ approaches
$\epsilon =0$. However, with numerical accuracy, it
merges with horizontal axis at $Q=0$. In Fig. \ref{fenergyplot}b the dimensionless
DOS, $\frac{dQ}{d\epsilon}$, is plotted for the branch $n=1$. This DOS contains two contributions from positive and from negative $Q$.
It is seen that the first contribution dominates for dimensionless energies $\epsilon>\epsilon_m$,
where $\epsilon_m$ is the minimal energy of the branch, see Fig. \ref{fenergyplot}a.

To find the energy dependence of the DOS in the domain $\epsilon\gg 1$,
we can use the asymptote (\ref{asymptote2}) and cast it into the form
\begin{equation}
\label{dispersion}
Q=\frac{\epsilon^4}{16n^2}.
\end{equation}
In  dimensional   units
Eq. (\ref{dispersion}) has the form
\begin{equation}
\label{dispersion}
k_x=\frac{y_0^3 E^4}{16 n^2v_F^4}.
\end{equation}
%We see that $\frac{dQ}{d\epsilon}\propto \epsilon^3$.
It is important that $\frac{dk_x}{dE}$
yields a {\em one-dimensional} DOS. To recalculate it into a real
%surface
2D DOS, we have to divide
$\frac{dk_x}{dE}$ by $R_c$.
%and return to the dimensional units
This leads to the result
\begin{equation}
\label{rhon}
\rho_{sn}^{(n)}(E)=\frac{1}{R_c}\frac{d k_x}{d E}=\frac{y_0^3E^3}{4 n^2R_c v_F^4}=\frac{\ell_B^2 E^3}{4n^2v_F^4}.
\end{equation}
Now it is easy to realize that,
upon summation over $n$, Eq. (\ref{rhon}) restores, within a factor,
the behavior $\phi(z)\propto z^3$ derived above
without accounting for the snake states.
From this we conclude that the true behavior of $\phi(z)$ at small $z$ is the one shown schematically
in the inset of Fig.~\ref{fenergyplot}b.
Namely, the $z^3$-dependence extends down to minimal $z=z_m$
and saturates below $z=z_m$. The value of $z_m$
can be found from Eq. (\ref{units})
\begin{equation}
\label{zm}
z_m=\frac{1}{w}\left(\frac{v_F}{y_0}\right)\sim\left(\frac{\ell_B}{R_c}\right)^{1/3}.
\end{equation}

%
%Returning to the dimensional units, the dispersion law Eq. (\ref{asymptote1})
%can be rewritten as
%\begin{equation}
%\label{dimensional}
%\Big(\frac{E}{w}\Big)^4=4n^2k_x\left(\ell_B^2R_c\right)^{1/3}.
%\end{equation}

%
%\begin{equation}
%\frac{1}{R_c}\frac{d k_x}{d E}=\frac{y_0^3E^3}{4 n^2R_c v_F^4}=\frac{\ell_B^2 E^3}{4n^2v_F^4}=z^3.
%\end{equation}

%The fact that extension of the electron wave function in the $y$ direction is much smaller than $R_c$

\section{Discussion}
\begin{figure}
\includegraphics[scale=0.45]{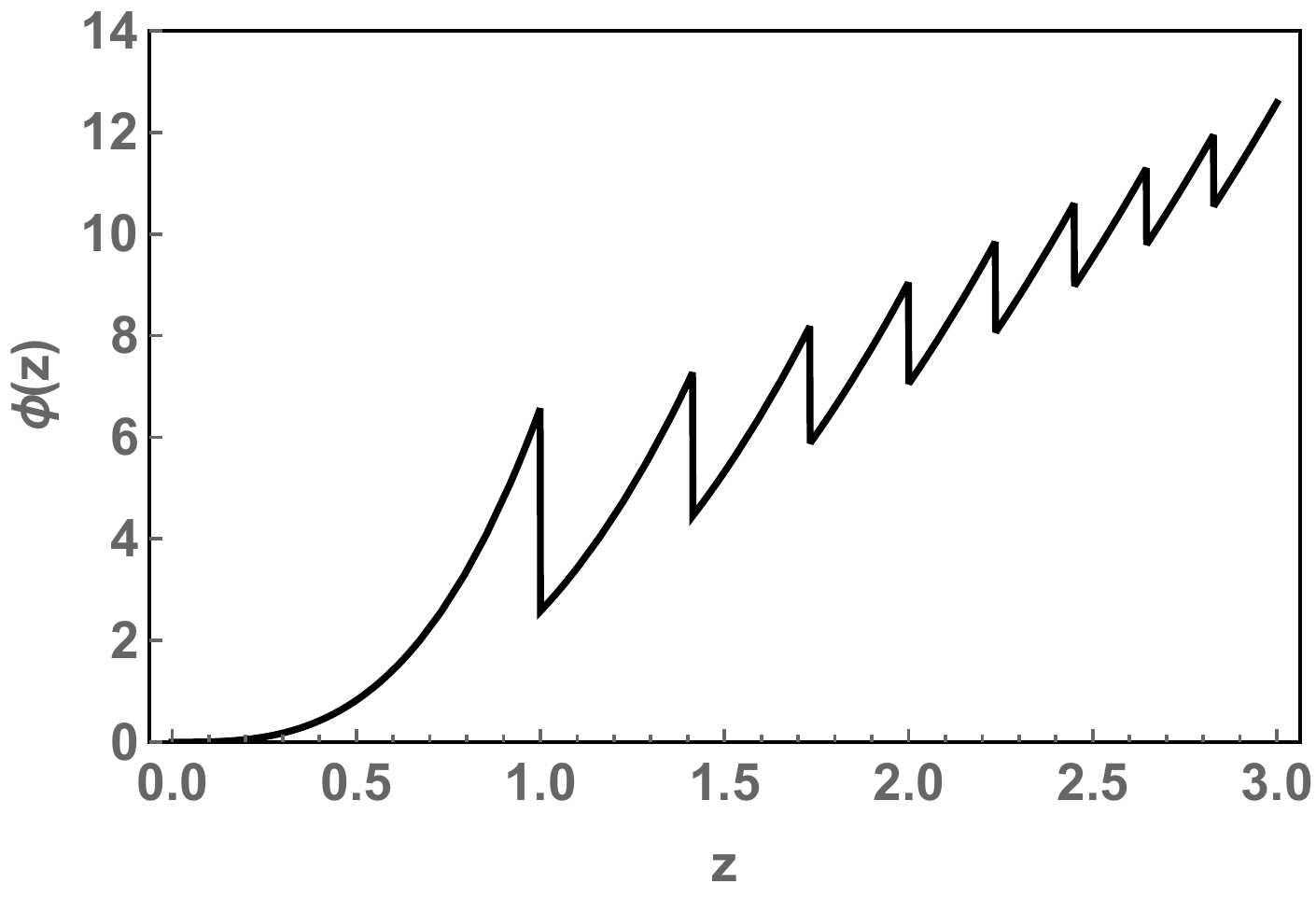}
\caption{(Color online) Average DOS is plotted from Eq. (\ref{averaged}) versus the dimensionless
energy $z=E/E_{{\text max}}$ for the homogeneous distribution of pseudo-magnetic fields
in the interval $\left(-B_{{\text max}},B_{{\text max}}\right)$. The energy $E_{{\text max}}$
is the Landau quantization energy in the field  $B_{{\text max}}$. Multiple features originate
from the sharp boundaries of the distribution. These boundaries are generic for the periodic arrangement of
ripples.}
\label{fsinosoidal}
\end{figure}
({\em i}) With regard to the observables, a pronounced peak in Fig. \ref{fDOS1}b
should manifest itself in the derivative of the {\em average}
differential conductance with respect to the bias voltage.
In other words, not only the zero-bias peak, seen in differential conductance\cite{BerkleyBubble}, but also $n=\pm 1$ peaks survive
the averaging over the ripples
due to the substrate topography.

A peak in the average ac conductivity, $\sigma(\Omega)$,
in Fig.~\ref{fconductivity} might manifest itself in the infrared characteristics
of the rippled graphene sample.
For example, for a typical field $1$T, the
maximum in $\sigma(\Omega)$ corresponds to the frequency $\Omega=120$K
and to the wavelength $100$$\mu m$.
%$\Omega=25$THz.
This frequency can be probed by photoconductivity\cite{Geim} or
by the measurement of the infrared transmission.\cite{Stormer}
In the latter case, the shape of transmission profile is determined by
$\frac{d\sigma\left(\Omega\right)}{d\Omega}$.
%\left(\Omega\right)$.

({\em ii}) From the STM images of graphene on different substrates,\cite{HeinzSTMimage,FuhrerTopography1,TopographySiO2,FuhrerTopographySi02,Zhitenev} it can be concluded that the realistic profile, $h(x,y)$, resembles a periodic
structure. Such a quasi-periodicity implies that the distribution of $B({\bf r})$ has a sharp cutoff at some $B=B_{{\text max}}$. This is because  $B({\bf r})$  is, essentially, the radius of curvature at the point ${\bf r}$.\cite{Maria,Lewenkopf1}
The consequence of this cutoff is a presence of numerous features in the average DOS corresponding to the Landau
levels in the field $B=B_{{\text max}}$. To illustrate this point, we chose a homogeneous distribution of pseudo-magnetic fields which translates into
$F(z)=\frac{1}{2}\theta(1-|z|)$.  The resulting DOS calculated
from Eq. (\ref{form}) is shown in Fig. \ref{fsinosoidal}. We see that $\phi(z)$
exhibits a sawtooth structure as it approaches the bare DOS.
More specifically, a microscopic treatment\cite{Maria,Lewenkopf1} of pseudo-magnetic fields induced by a bubble of concrete shape and with
angular symmetry indicates that the maximum of pseudo-magnetic field corresponds
to the points away from the center. Then the distribution, $F(B)$, diverges near
$B_{{\text max}}$ as $(B_{{\text max}}-|B|)^{-1/2}$.

({\em iii})
We emphasize that the inflection point in Fig. \ref{fDOS1} is the consequence
of Landau levels in the Dirac spectrum being non-equidistant. If they were
equidistant, the function $\phi(z)$ would take the form 
$\phi(z)=z\sum_n\frac{1}{n^2}\exp\left(-\frac{z^2}{n^2}\right)$ instead of Eq. (\ref{form}).
Plotting this function indicates that the feature near $z=1$ is pronounced very weakly.

({\em iv})
%\begin{equation}
%A_x({\bf r})=\nu\frac{\Phi_0}{a}\left[u_{xx}({\bf r})-u_{yy}({\bf r})\right],~~
%A_y({\bf r})=-2\nu\frac{\Phi_0}{a}u_{xy}(\bf{r}),
%\end{equation}
To explore the applicability of the adiabatic description
%To perform the numerical estimate,
we use the expression\cite{ExpressionForB}
for pseudo-vector potential in terms of
the surface profile,
$h(\bf r)$,
\begin{eqnarray}
\label{vectorpotential}
A_x({\bf r})&=&\nu\frac{\Phi_0}{a}\Bigg[\left(\frac{\partial h}{\partial x}({\bf r})\right)^2-\left(\frac{\partial h}{\partial y}({\bf r})  \right)^2\Bigg], \nonumber\\
A_y({\bf r})&=&-2\nu\frac{\Phi_{0}}{a}\Bigg(\frac{\partial h}{\partial x}({\bf r})\Bigg)\Bigg(\frac{\partial h}{\partial y}({\bf r})\Bigg),
\end{eqnarray}
where $\Phi_0=\frac{\hbar c}{e}$ is the flux quantum,
and $a=1.42{\AA}$  is the lattice constant.
Parameter $\nu\approx 0.6$ is expressed via
the rate of change of the overlap intergral\cite{ExpressionForB}
with $a$. From  Eq. (\ref{vectorpotential})
we find the following general expression for the
pseudo-magnetic field
\begin{equation}
\label{ExpressionForB}
B({\bf r})=2\nu\frac{\Phi_0}{a}\left[2\frac{\partial h}{\partial x}\frac{\partial^2h}{{\partial x}{\partial y} }-\frac{\partial h}{\partial y}\left(\frac{\partial^2h}{\partial y^2}-\frac{\partial^2h}{\partial x^2}\right)\right].
\end{equation}
To calculate the average $\langle B^2({\bf r})\rangle$ it is convenient to switch in Eq. (\ref {ExpressionForB}) to the Fourier transform, $h({\bm r})=\sum_{\bm q}h_{\bm q}\exp\left(i{\bm q}\cdot{\bm r}\right)$
\begin{multline}
\label{Bsquare}
B^2({\bf r})=
4\nu^2\frac{\Phi_0^2}{a^2}|h_{\bm q}|^2|h_{\bm Q}|^2\\
\times \Big[4(q_x^2-q_y^2)Q_x^2Q_y^2+4q_x q_y Q_x Q_y(Q_x^2-Q_y^2)+q_y^2(Q_x^2+Q_y^2)^2   \Big].
\end{multline}
Only the last term contributes to $\langle B^2({\bf r})\rangle$.
For the Gaussian correlator,
\begin{equation}
\label{Correlator}
\langle h({\bf r}_1)h({\bf r}_2)\rangle=
h_0^2\exp\left[-\frac{\left({\bf r}_1-{\bf r}_2\right)^2}{R_c^2}\right],
\end{equation}
evaluation of Eq. (\ref{Bsquare}) yields
%\begin{equation}
%u_{xx}=\frac{1}{2}\left(\frac{\partial h}{\partial x}\right)^2,~~u_{yy}=\frac{1}{2}\left(\frac{\partial h}{\partial y}\right)^2,~~~
%u_{xy}=\frac{1}{2}\left(\frac{\partial h}{\partial x}\frac{\partial h}{\partial y}\right )
%\end{equation}

\begin{equation}
\label{BsquareFinal}
\langle B^2({\bf r})\rangle=
\left(\frac{16\nu\Phi_0h_0^2}{R_c^3a}\right)^2.
%\sum_{\bm q}{\bm q}^2|h_{\bm q}|^2\sum_{\bm Q}{\bm Q}^4|h_{\bm Q}|^2.
\end{equation}
The latter expression can be rewritten in the form
\begin{equation}
\label{ratio}
\frac{\ell_B}{R_c}=\left(\frac{R_ca}{16\nu h_0^2}\right)^{1/2}.
\end{equation}
For adiabatic description to apply the right-hand side should be
small. It requires strong corrugation amplitude $h_0$, while, somewhat counterintuitively,
the this corrugation cannot be too smooth for Landau levels to form within $R_c$.
(the case opposite to Ref. \onlinecite{OneBubbleNancySandler}).
Using the parameters $R_c=4$nm and $h_0=0.5$nm from Ref. \onlinecite{BerkleyBubble} we
get for the right-hand side the value $0.47$. Similar parameters $R_c=4\pm 2$nm and
$h_0=0.6\pm 0.1$ were extracted in Ref. \onlinecite{Folk2} from the transport measurements.
Concerning the STM measurements, 
the parameters $R_c\approx 10$nm and $h_0\approx 0.5$nm reported in different papers 
\onlinecite{HeinzSTMimage,FuhrerTopography1,TopographySiO2,FuhrerTopographySi02} are consistent
with each other and correspond to the value $0.74$ in the right-hand side.

%\begin{equation}
%F(B)=\delta\big(B-\sum_{{\bf q}_1,{\bf q}_2}h_{{\bf q}_1} h_{{\bf q}_2}W({\bf q}_1,{\bf q}_2)\big)
%\end{equation}
%
%
%
%
%({\em iv}) Criterion for adiabatic description.
%Our crucial  assumption is that
%
%Pseudo-magnetic field is a quadratic function of $h(x,y)$. Why the gaussian distribution\cite{Lewenkopf,Morpurgo} of $h(x,y)$  leads to the gaussian distribution of pseudo-magnetic field?
%

%4. Features in the DOS is a consequence of fact that the Landau levels in the Dirac spectrum being non-equidistant.

\vspace{2mm}

\section{Acknowledgements}

\vspace{2mm}
We are grateful to R. K. Morris who participated in this work at the early stage.
The work was supported by the Department of Energy,
Office of Basic Energy Sciences, Grant No.  DE- FG02-
06ER46313.

\end{document}